 \documentclass[preprint]{aastex}

\slugcomment{To appear in ApJL}

\begin{document}

\title{The distance to the Fornax dwarf galaxy using red clump stars, and the
discrepancy between red clump and tip of the red giant branch distances}

\author{D. Bersier}
\affil{Harvard-Smithsonian Center for Astrophysics, 60 Garden St., Cambridge MA 02138}
\email{dbersier@cfa.harvard.edu}

\begin{abstract}

I determine a distance to the Fornax dwarf galaxy using stars in the red clump
and at the tip of the red giant branch. They are in very good agreement,
with $\mu_0 = 20.66\ mag$. Comparing the
magnitudes of the tip of the red giant branch and of the red clump in Fornax,
Carina and the Magellanic Clouds, I propose a possible solution to the problem
of the discrepancy between these two  types of distance measurements.

\end{abstract}


\keywords{galaxies: distances and redshifts --- galaxies: individual (Fornax,
Large Magellanic Cloud, Carina) --- stars: late-type}

\section{Introduction}
\label{sec_intro}

In the recent years, there has been a flurry of activity about the use of
red clump stars as a distance indicator. Several authors provided examples of
distances determinations with the RC method to the Magellanic Clouds, the Galactic
Center, the Carina dwarf galaxy and M31 (e.g. Paczy\'nski \& Stanek 1998; Stanek et al.
1998; Stanek \& Garnavich 1998; Udalski 1998, Udalski 2000).
There are several strong arguments in favor of this method. First, the red clump is easy
to recognize in the color-magnitude diagram and this feature is little affected
by other structures in the CMD (like the horizontal branch, the red giant branch
or foreground contamination). Second, there are many red clump stars in a typical
galaxy or cluster, this renders the method statistically robust.
Third, it is one of very few distance indicators that can potentially be accurately
calibrated, with several hundreds of stars having Hipparcos parallaxes better than 10\%
\citep{ps98}.
However, there are complications. Several authors (Girardi et al. 1998; Girardi 1999;
Cole 1998) presented arguments, mostly theoretical, that the average magnitude
of the red clump is sensitive to age and metallicity and thus, a detailed
knowledge of the average age and metal content of a galaxy are needed before using
the RC method to determine distances.
The debate has been heated by the fact that the advocates of the RC method found very
short distances to the Magellanic Clouds, e.g. $\mu_{LMC} = 18.07\ mag$ in
\citet{szh98} and
$\mu_{LMC} = 18.08\ mag$ in \citet{uda98}, later revised upwards
to $\mu_{LMC} = 18.24\ mag$
\citep{uda00}. The reddening correction in the LMC may also be
a source of disagreement since the average red clump magnitude measured by different
groups -- using different methods to correct for extinction -- are markedly different
[$I_{0,m} = 17.84\ mag$ in \citet{szh98}, $I_{0,m} = 18.06\ mag$ in \citet{szk00},
$I_{0,m} = 18.12\ mag$ in \citet{rom00}].

Most recently, \citet{uda00} and \citet{pop00} calibrated empirically
the metallicity dependence of $I_{0,m}$. Their slopes and zero points are in agreement
(within $1\sigma$ of each other). For instance, \citet{pop00} obtained
\begin{equation}
\label{eq_mi1}
M_I^{RC} = 0.19\mbox{[Fe/H]} - 0.23 \ mag
\end{equation}
while Udalski obtained
\begin{equation}
\label{eq_mi2}
M_I^{RC} = 0.13\mbox{[Fe/H]} - 0.23 \ mag
\end{equation}
Accounting for this correction, these authors obtain a distance modulus to the
LMC of $18.27\ mag$ and $18.24\ mag$ respectively.

The aim of this letter is twofold. First, to provide a test of the accuracy of the
RC method, and more particularly of the metallicity correction.
Second, to explore the influence of reddening corrections, by working
in very low reddening systems (Fornax and Carina).
I also compare red clump distances for four galaxies (Carina, Fornax and the
Magellanic Clouds) with distances obtained via the tip of the red giant branch and
discuss the consequences.

\section{Observations and data reduction}
\label{sec_obs}

About 30 epochs of photometry have been obtained on two telescopes with the aim
of searching for short period variable stars in the Fornax dwarf galaxy (Bersier
\& Wood, 1999, 2000). Here I use only $\sim$ half the dataset that has been
calibrated. Full details on the data reduction will appear elsewhere (Bersier \&
Wood, in preparation), only a short summary is given here.
Four fields in Fornax have been observed between October and December 1997 with the
40\arcsec\ telescope at Siding Spring Observatory (Australia), using $V$ and $I_C$
filters. The detector was a SITe $2048 \times 2048$ CCD that gave a field of
view of $20\arcmin\times
20\arcmin$. Between 12 and 18 images in $V$ and $I$ have been obtained for each of the
four fields. The photometry has been done with DoPHOT \citep{sms93} and calibrated
using observations of Landolt standard stars
\citep{lan92}. From observations of stars in overlapping fields, I estimate that
the absolute photometric zero-point is good to $\pm 0.02\ mag$ for each passband.

\begin{figure}
\epsscale{0.8}
\plotone{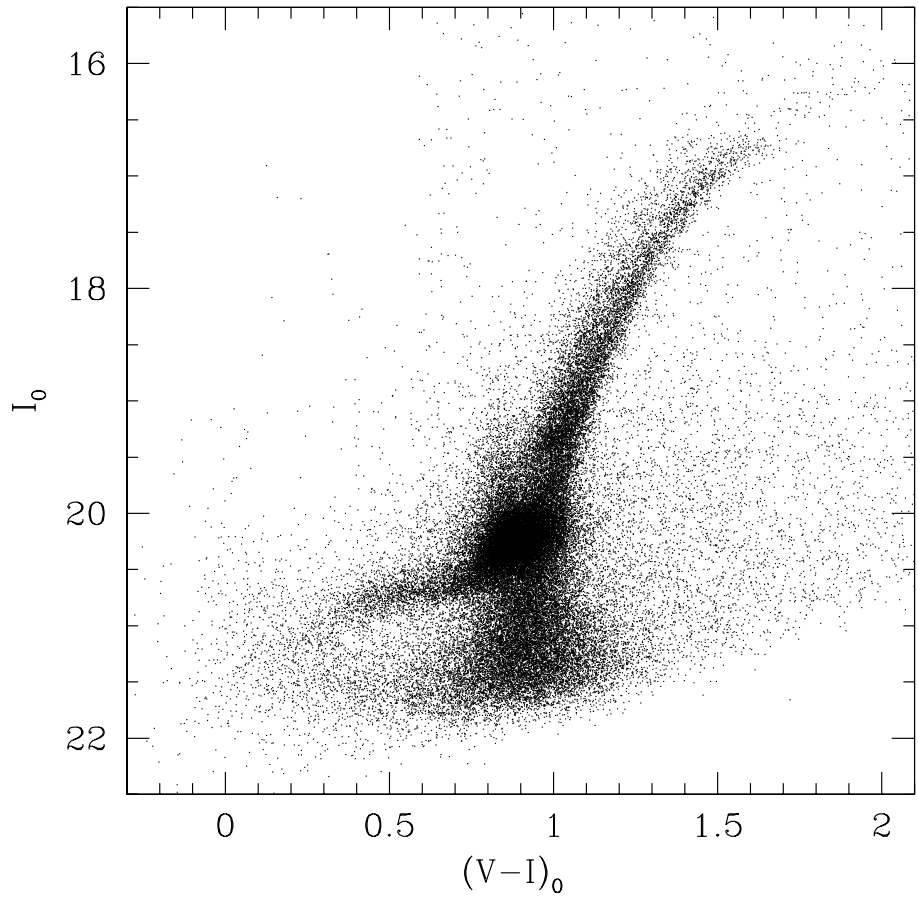}
\caption{
The $I_0,\, (V-I)_0$ color-magnitude diagram for the observed
field ($\sim 40'\times 40'$). Almost 60~\% of the stars are in the red clump.
\label{fig_cmd}}
\end{figure}

Since there is a large number of observations per field, I computed the average of
all $VI$ measurements for each star. This clearly improves the photometric
accuracy; the average error at the magnitude of the red clump ($m_I\simeq 20.3$)
is $\sim 0.1$ mag.
I then used the maps of \citet{sfd98} to correct for reddening. This yielded values
of $E(B-V)$ between 0.02 and 0.04 for the vast majority of the stars. One then
has $E(V-I) = 1.28 E(B-V)$ and $A_I = 1.96 E(B-V)$.
The color-magnitude diagram for the $40\arcmin\times 40\arcmin$
field is shown on Fig.~\ref{fig_cmd}.

\section{The distance to Fornax}

\subsection{The red clump}
\label{sec_rc}

The red clump appears to be very compact and well limited in color between
$(V-I)_0\simeq 0.8$ and $(V-I)_0\simeq 1.0$ (see Fig.~\ref{fig_cmd}). To obtain
a distance, the first step is to fit a function of the type 
\begin{eqnarray}
N(I_0) & = & a + b\,(I_0 - I_{0, m}) + c\,(I_0 - I_{0,m})^2  \nonumber \\
 & & + \frac{N_{RC}}{\sigma_{RC}\sqrt{2\pi}}\, \exp\left[ - \frac{(I_0 - I_{0,m})^2}{2\sigma^2_{RC}}  \right]
\label{eq_rcfit}
\end{eqnarray}
following \citet{ps98}. The Gaussian  represents the red clump itself and the
parabola accounts for the ``background'' giants.
Usually, stars are selected in the color range
$0.8< (V-I)_0 < 1.25$. It seems that in Fornax the red clump reaches slightly bluer
colors than 0.8, however this is also where the fainter part of the clump merges
with the horizontal branch.
From the fits performed for different color ranges, it is clear that bluer than
$(V-I)_0=0.8$ the red clump is significantly fainter than for $(V-I)_0>0.8$.
The fact that the width of the clump is larger for $(V-I)_0<0.8$ than for $(V-I)_0>0.8$
is also an indication that the clump is contaminated by the horizontal branch.

\begin{figure}
\epsscale{0.8}
\plotone{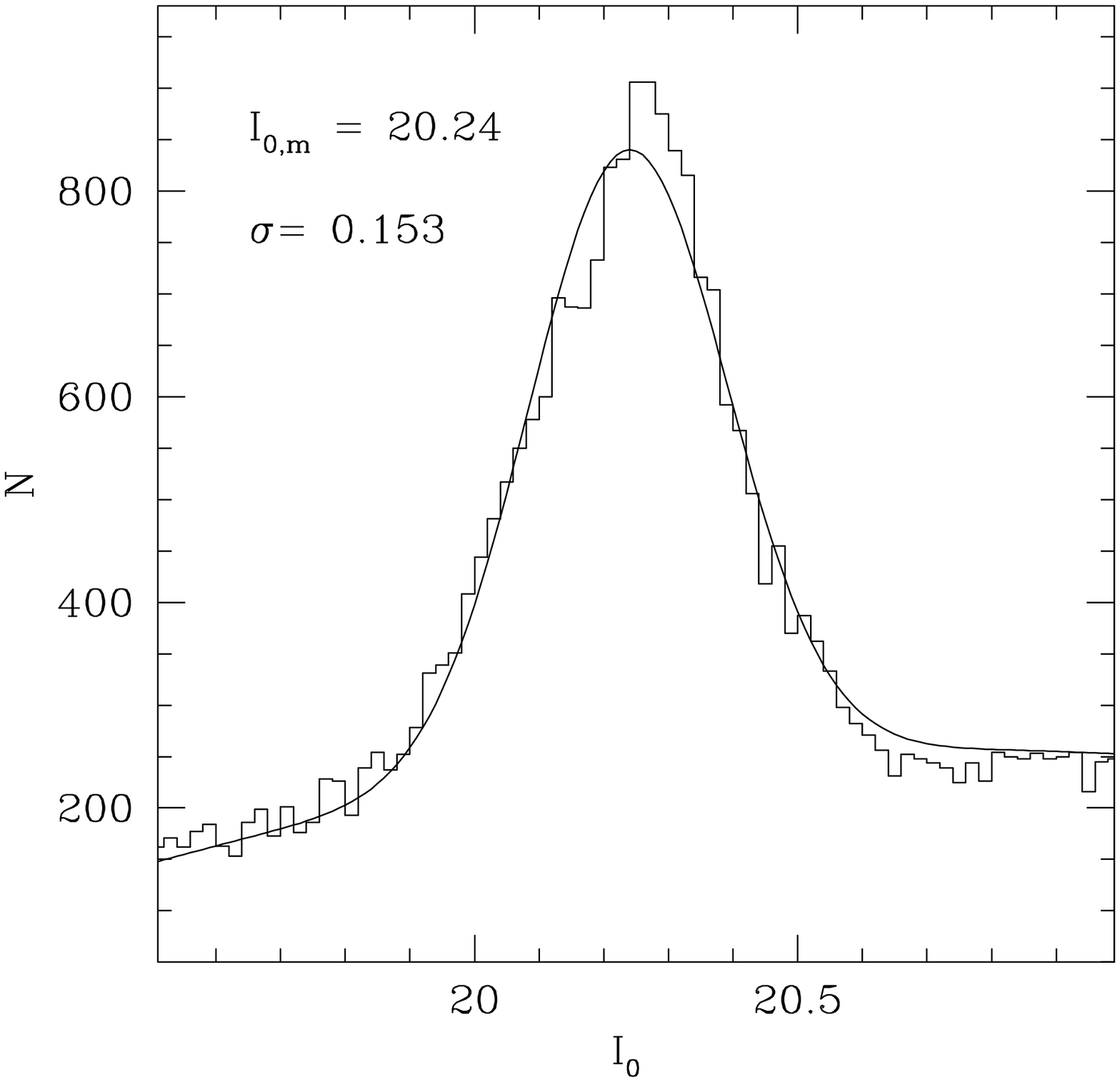}
\caption{ Magnitude distribution of red clump stars with colors in
the range $0.8 < (V-I)_0 < 1.25$. The solid line is the fit of Eq.~1.
\label{fig_fit}}
\end{figure}

Fitting eq.~\ref{eq_rcfit} to the whole red clump sample ($0.7 - 1.25$), one
obtains $I_{0,m} = 20.25 \pm 0.004$.
Restricting the color range to $0.8\leq (V-I)_0 \leq 1.25$ gives
$I_{0,m} = 20.24 \pm 0.004$ (see Fig.~\ref{fig_fit}).

In order to obtain a distance, a metallicity correction to $I_{0,m}$ needs to be
applied. I took [Fe/H]$_{RC} = -1.0$ for two reasons: {\it i)} The average
color of Fornax's red
clump is intermediate between the [Fe/H] $ = -1.3$ and [Fe/H] $ = -0.7$ cases of
\citet{gir99a}; {\it ii)} \citet{shb00} recently argued that the average metallicity
of Fornax is $-1.0$.  Using Popowski's relation \citep{pop00} yields $M_I^{RC} = -0.42$,
then $\mu^{RC}_0 = 20.66$.; using Udalski' relation, one has $\mu^{RC}_0 = 20.60$.

\subsection{The tip of the red giant branch}
\label{sec_trgb}

It has been shown that the tip of the red giant branch is a reliable and accurate
distance indicator (e.g. Lee et al. 1993; Sakai et al. 2000 and references therein).
Comparisons between TRGB distances and Cepheid distances in a number of galaxies
have shown that there is excellent agreement between these two distance indicators
(e.g. Lee et al. 1993, Kennicutt et al. 1998, Ferrarese et al. 2000).
The absolute magnitude of the tip of the red giant branch (TRGB) is almost
constant, $M_I^{TRGB} \simeq -4.0 \pm 0.1$  for [Fe/H]$\lesssim -0.7$ \citep{lfm93}.

\begin{figure}
\epsscale{0.8}
\plotone{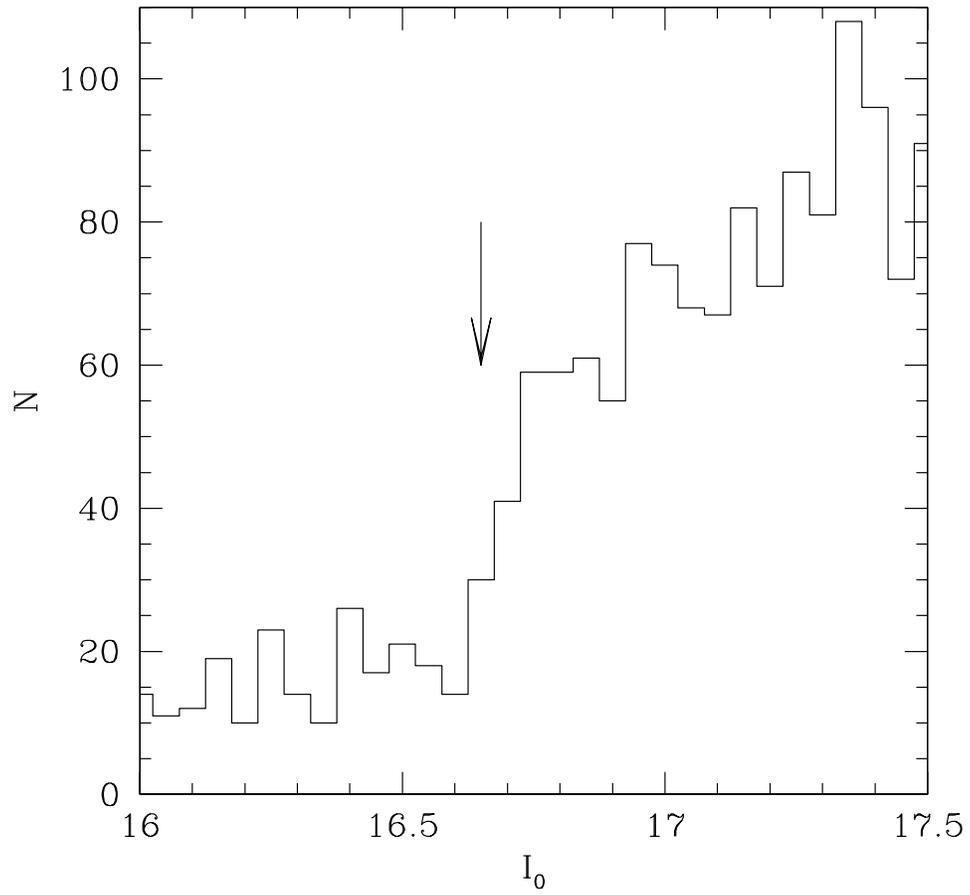}
\caption{
The $I$-band luminosity function for bright stars. The tip of
the red giant branch is identified by the arrow at $I_0 = 16.65$.
\label{fig_trgb}}
\end{figure}

The average age and metallicity of red giants in Fornax are such that the TRGB
method can safely be used.
The $I$-band luminosity function of Fornax is shown on Fig.~\ref{fig_trgb}. The tip
of the RGB is identified as the sharp number increase at $I_0 = 16.65$; the error is
estimated to 0.05 mag. This yields a distance modulus of
$\mu_0 = 20.65 \pm 0.11$ mag. It agrees very well with the distance modulus
derived above with the red clump. It is also in perfect agreement with the distance
derived by \citet{shb00} with the same method.

\section{Discussion}

The red clump and TRGB distances agree in Fornax, it is thus puzzling that
they don't agree in the LMC (Romaniello et al. 2000; Sakai et al. 2000). It might be
worth looking at other galaxies with RC and TRGB distances to see how these distances
compare. The difference
$\Delta I = I_{RC} - I_{TRGB}$ does not depend on a particular calibration of the
distance scale, it is an observable quantity. Moreover, it should not be sensitive to
reddening correction, hence it can be used to compare the two methods.
I collected data in the literature for Carina, the Small and Large Magellanic Clouds.
For Carina I took the red clump magnitude
$I_m$ from \citet{uda98} and the TRGB magnitude from \citet{ssh94}. These authors used
different values for the reddening $E(B-V)$ so I used the observed magnitudes $I_m$
and $I_{TRGB}$ \emph{without} reddening correction; these magnitudes are given in
Table~\ref{table_mags}. For the SMC, I used the data made publicly available by
the OGLE team \citep{ogle98}. For the LMC I used the data of \citet{zht97}.
I also took the TRGB and RC magnitudes (corrected for reddening) published in
\citet{szk00} for the LMC.
%All red clump magnitudes have been corrected according to Eq~\ref{eq_mi1}.
All these numbers as well as $\Delta I$ are given in
Table~\ref{table_mags}.

\begin{figure}
\epsscale{0.8}
\plotone{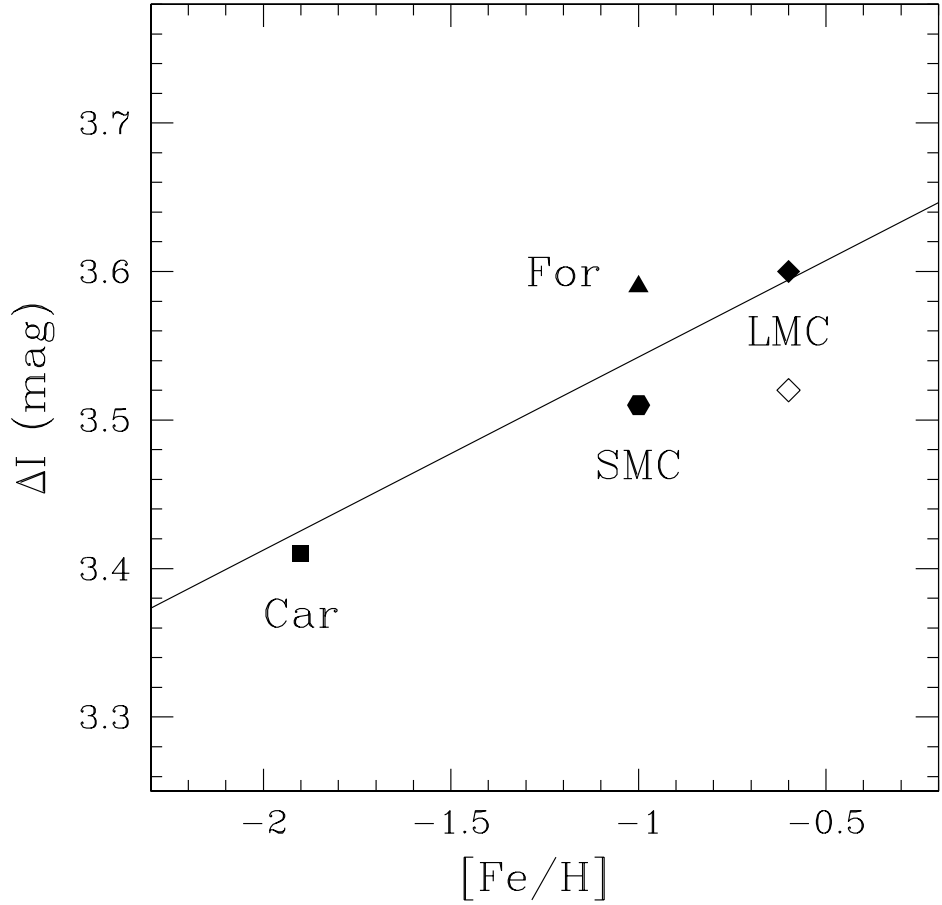}
\caption{ The magnitude difference $\Delta I = I_{RC} - I_{TRGB}$
plotted as a function of metal content [Fe/H]. Each galaxy name is indicated.
The solid line is the difference in
absolute magnitudes when using Udalski's relation for the absolute magnitude of the
red clump and with $I_{TRGB} = -3.9$. Filled symbols indicate that no reddening
correction has been applied, the empty symbol indicate that a reddening
correction has been applied. Error bars have been ommited for clarity.
\label{fig_deltami}}
\end{figure}

The TRGB is calibrated with Galactic globular clusters that are
old and metal-poor systems \citep{dca90}. The tip of the RGB is thus calibrated
for ages  larger than $\gtrsim 2$ Gyr and
[Fe/H]$\lesssim -0.7$. We know that these conditions are met in Carina and Fornax
since these galaxies contain old and metal-poor stars. It is also well known that
the LMC has a population of old and metal-poor stars (as attested by the many thousands
of RR Lyrae known in the LMC). However, stars of intermediate ages (2-3 Gyr) and/or
more metal-rich than [Fe/H]$=-0.7$ do exist in the LMC and they could contaminate
the TRGB. In other words, it could be that the TRGB detected by \citet{szk00}
is that of a young population. This seems very unlikely in view of the $I$-band
luminosity function they present. The only possible brighter discontinuity in
their luminosity function could be at $\sim 14.2$ (although it looks more like a
noise spike).
Furthermore, metal-rich stars younger than $\sim 3$ Gyr would be distributed over
a fairly wide range in
magnitude and they would smooth out the $I-$band luminosity function, they would
not create such a sharp discontinuity. Hence the magnitude $I^{TRGB} = 14.54$ (from
Sakai et al. 2000) must correspond to the tip of the red giant
branch of the old population, and it is not affected by
intermediate age stars. Only very young red supergiants are as bright or brighter
than old red giants, however there are very few supergiants compared to the number
of RGB stars. Hence the TRGB magnitude should be robust, even in a galaxy that
contains young and intermediate age populations such as the Magellanic Clouds.

The magnitude difference $\Delta I$ is plotted as a function of [Fe/H] in
Fig.~\ref{fig_deltami}. The solid line in this figure is the magnitude difference
obtained with $I_{TRGB} = -3.9$ mag and $I_{RC} = 0.13 [Fe/H] - 0.23$ \citep{uda00}.
Note that in this figure all points except the lower LMC point have not been corrected
for reddening.
Two facts deserve some comments.
First, there is a disagreement between the two LMC measurements. Both are based
on the same data set \citep{zht97} but one has been corrected for extinction while
the other has not. In his referee report, A. Udalski used OGLE2 data
to determine $\Delta I$ \emph{after} correcting for reddening, using the OGLE
reddening map \citep{uda99}. He finds $\Delta I = 3.62$, virtually identical to
3.60 found here without reddening correction. This shows that the source of the
discrepancy between TRGB and RC distance measurements probably lies in the
extinction correction rather than in the intrinsic properties of one or the other
of these distance indicators.

Second, the solid line seems to represent very well the behavior of the difference
$I_{RC} - I_{TRGB}$, provided that the same reddening correction is applied to
both red clump stars and tip giants, assuming that Udalski's relation correctly
predicts the metallicity dependence of the red clump and that the TRGB absolute
magnitude is $I_{TRGB} = -3.9$ mag. Note that using Popowski's relation for the red
clump \citep{pop00} would change $\Delta I$ by a few hundredths of a magnitude only.
The TRGB is calibrated using RR Lyrae stars \citep{dca90}; the recent revision
of RR Lyrae magnitudes advocated by \citet{pg98} would decrease the brightness of
the TRGB. If the above procedure is correct, these two distance indicators can be in
much better agreement than previously thought.


\acknowledgments

I thank Kris Stanek for helpful discussions and Dennis Zaritsky for putting
his MCPS data at my disposal. I also thank the referee, A. Udalski,
for his challenging comments that improved the presentation and content of this paper.
This work has been partially supported by NSF grant AST-9979812 and by the Swiss NSF
(grant 8220-050332).


\clearpage

%%   Tables

\begin{table}
\begin{center}
\caption{Values of the average clump magnitude \label{table_fits}}
\begin{tabular}{ c c c l }
\tableline\tableline
$(V-I)_0$ & $I_{0,m}$\tablenotemark{a} & $\sigma_{RC}$\tablenotemark{b} & Comments \\
color range & & &  \\
\tableline
%%  for the lines below, all in the mag range 19.5 - 21.0
0.8 -- 1.25 & 20.24 & 0.153 & \\
0.8 -- 1.25 & 20.30 & 0.158 & No reddening correction\\
0.8 -- 1.0  & 20.25 & 0.155 & \\
0.7 -- 1.25 & 20.25 & 0.164 & Affected by HB \\
\tableline
\end{tabular}
\tablenotetext{a}{For all fits the magnitude range was 19.5 to 21.0.}
\tablenotetext{b}{$\sigma_{RC}$ is the width of the Gaussian (see eq.~1).}
\end{center}
\end{table}

\clearpage

\begin{table}
\begin{center}

\caption{The magnitude difference between the tip of the red giant branch and
the red clump for several galaxies  \label{table_mags}}
\begin{tabular}{ l c c c c c}
\tableline\tableline
Galaxy & Car & For & SMC\tablenotemark{a} & LMC\tablenotemark{b} & LMC\tablenotemark{c} \\
\tableline
\mbox{[Fe/H]\tablenotemark{d}} & $-1.9$ & $-1.0$ & $-1.0$ & $-0.6$ & $-0.6$ \\
$I_{TRGB}$\tablenotemark{e} & 16.15 & 16.65 & 15.00 & 14.60 & 14.54 \\
$I_{RC}$\tablenotemark{f}   & 19.56 & 20.24 & 18.51 & 18.20 & 18.06 \\
$\Delta I$\tablenotemark{g} & 3.41 & 3.59 & 3.51 & 3.60 & 3.52 \\
%Age (Gyr) & $\sim 5$ & 5 & 1.5 \\
\tableline
\tableline
\end{tabular}
\tablenotetext{a}{Magnitudes in this column are based on OGLE data \citep{ogle98}}
\tablenotetext{b}{Magnitudes in this column are based on data from \citet{zht97}}
\tablenotetext{c}{Magnitudes in this column are taken from \citet{szk00}}
\tablenotetext{d}{The [Fe/H] values are from \citet{ssh94} and \citet{uda98} for Carina;
\citet{shb00} and this paper for Fornax; \citet{uda98} for the LMC and SMC.}
\tablenotetext{e}{$I_{TRGB}$ is the observed magnitude of the tip of the RGB}
\tablenotetext{f}{$I_{RC}$ is the observed average magnitude of the red clump}
\tablenotetext{g}{$\Delta I = I_{RC} - I_{TRGB}$}
\tablecomments{Uncertainties on $I_{TRGB}$ are $\sim 0.05$ mag, on $I_{RC}$ they are
of order 0.03 mag.}

\end{center}
\end{table}

\end{document}